\documentclass[twocolumn, pra]{revtex4-1}

\usepackage{booktabs}
\usepackage{palatino}
\usepackage{verbatim}
\usepackage{amsmath,amsthm}
\usepackage{amssymb}
\usepackage{graphicx}
\usepackage{color}
\usepackage{tikz}
\usetikzlibrary{backgrounds,fit,decorations.pathreplacing,calc,,shadows}

\usepackage{natbib}
\usepackage{epsfig}

\def\ket#1{|{#1}\rangle}

\newcommand{\ketbra}[2]{|#1\rangle\langle#2|}

\newcommand{\trace}{\operatorname{Tr}}

\begin{document}


\title{Hamiltonian Composite Dynamics Can Almost Always Lead To Negative Reduced Dynamics}

\author{James M. McCracken}
\email{james.mccracken@nrl.navy.mil}
\affiliation{U.S. Naval Research Laboratory, Code 5540, Washington, DC 20375}
\date{\today}

\begin{abstract}
Complete positivity is a ubiquitous assumption in the study of quantum systems interacting with the environment.  It will be shown that Hamiltonian evolution of a quantum system and its environment can be negative (i.e.\ not completely positive) in the energy basis, by showing that such evolution is {\it almost always} negative for given initial conditions.  As such, ignoring or ``correcting'' experimental data that is not completely positive may cause the loss of important information regarding system-environment correlations and coupling.  Complete positivity assumptions are an important part of many quantum information theories, and it is important to understand how (and why) it appears to be possible to violate the complete positivity requirement in the examples presented here.  A relationship between the negativity of an evolution and the eigenvalues of the Hamiltonian will be shown, and experimental verification of negative reduced dynamics will be proposed.
\end{abstract}
\maketitle

\section{Introduction}
Complete positivity has become an ingrained part of the modern study of open quantum systems.  Quantum information channels are usually defined as ``completely positive (CP), trace preserving maps'' on quantum states \cite{nielsen00}.  Discussions of CP \footnote{``CP'' is overloaded to mean both ``completely positive'' and ``complete positivity''.  The use should be clear in context.} requirements and violations have appeared in the literature for two decades \cite{Pechukas1,Alicki95,Pechukas2,Sud05,Rodríguez07,Terno08} (and references therein), yet most modern open systems and quantum information textbooks state CP as a requirement for the evolution of a quantum system with little or no theoretical justification \cite{nielsen00,breuer07,alicki07,benatti03}.  This issue has become more prevalent as non-CP experimental evidence continues to appear in the literature \cite{howard06,Boulant03}.  Tomographic characterization of qubit channels are common quantum information experiments, and it will be shown below that tomography is closely tied to the concept of complete positivity.  It is possible to write down reduced dynamics that appear to violate the CP requirement yet still seem to have straightforward experimental interpretations.  If theoretical descriptions of experiments can be provided that violate the CP requirement, then it is important to understand if those theoretical descriptions are not physical for some reason, or if the CP assumption is not valid for all reduced dynamics.  The CP requirement may limit, a priori, the theoretical evolutions of a quantum system without empirical justification.

Many authors have argued that the reduced dynamics of quantum systems need not be CP (see the references listed above).  It will be shown here that evolution of a composite quantum system governed by any Hamiltonian is only CP under certain, very specific conditions.  A previous non-CP example exists in the literature \cite{Rodriguez2008}, but it will shown here that, given certain initial conditions, any Hamiltonian will almost always lead to negative dynamics in a given basis.

A non-CP quantum evolution is called ``negative''.  A composite quantum system is a quantum system under the control of the experimenter (called the ``reduced system'') along with the other quantum systems inaccessible to the experimenter that may still influence the dynamics of the reduced system (called the "bath", "environment", "reservoir", etc).  It will be shown that a measurement of the negativity (defined below) can give an experimenter some understanding of the coupling and correlations between the reduced system and bath.

\section{Reduced Dynamics}
A quantum channel is a map that takes density matrices to density matrices, i.e.\ $\varepsilon(\rho) = \rho(t)$, where $\rho$ is the state of the system of interest at time $t=0$ and $\rho(t)$ is the state of that system at some later time $t$.  This evolution is called the ``reduced dynamics'' in the open systems literature.  A quantum channel is defined as
\begin{equation}
\label{eq:channel}
\varepsilon(\rho) = \trace_B\left(U \rho^\sharp U^\dagger\right)\;\;,
\end{equation} 
where $\rho$ is the initial state of the reduced system, $U$ is the unitary evolution of the composite system, and $\sharp$ is called the ``assignment map'' (or ``sharp operator'').  The state $\rho$ ``resides'' in the Hilbert space accessible to the experimenter in the lab, $\mathcal{H}^S$, and the evolution of the reduced system is found by ``tracing out'' the bath from the joint evolution of the reduced system and the bath \cite{breuer07}; i.e.\ $U$ is associated with the composite Hilbert space $\mathcal{H}^{SB}=\mathcal{H}^S\otimes\mathcal{H}^B$ where $\mathcal{H}^B$ is the Hilbert space of the bath.  All Hilbert spaces in this work are assumed to be finite dimensional.  The partial trace operation, $\trace_B$, is an operator that allows expectation values of observables in the reduced system to be consistent with trivial extensions into a higher dimensional Hilbert space \cite{cohen77}.  

\begin{figure}[h!]   
\centering
\includegraphics[width=0.45\textwidth]{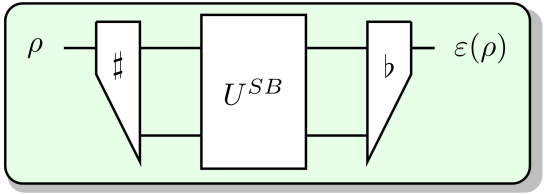}
\caption{This is an open circuit diagram for the channel described by Eqn.\ \ref{eq:channel}. The $\flat$ operator is defined as the partial trace with respect to the bath, i.e.\ $\rho^\flat\equiv\trace_B(\rho)$, and it is simply a convenient, space saving notation.}
\label{fig:diagram}
\end{figure}

Assignment maps were originally introduced by Pechukas in \cite{Pechukas1,Pechukas2} and were studied further in \cite{rodriguez10}.  The assignment map is an operation that injects the initial state of the reduced system into the higher dimensional Hilbert space of the composite system (i.e.\ $\rho^\sharp$ ``resides'' in $\mathcal{H}^{SB}$).  Channels are characterized in the lab through tomography, hence $\varepsilon$ needs to be linear, which implies the $\sharp$ operator is also linear.  The channel should take valid quantum states to valid quantum states, hence $\varepsilon$ needs to be positive (on some domain of states) and hermiticity-preserving.  Ergo, $\sharp$ should be positive (on some domain of states) and hermiticity-preserving.  Finally, $\sharp$ needs to be consistent, i.e.\ $\trace_B\left(\rho^\sharp\right) = \rho$.  A diagram showing the schematic behaviour of the assignment map and the partial trace is shown in Fig.\  \ref{fig:diagram}.      

The sharp operator describes a preparation procedure (as will be shown later), but the exact physical interpretation of the sharp operator is not clear.  Several different composite states can give rise to the same reduced state (e.g.\ if the state of the composite system is a maximally entangled pair, then the state of the reduced system will always be the completely mixed state independent of the specific maximally entangled state of the composite system).  The sharp operator simply gives one possible composite state for each reduced state.  It is well defined mathematically because of the mathematical restrictions desired for the channel $\varepsilon$.  It was shown by Pechukas \cite{Pechukas2} that the only assignment map that is consistent, linear, and positive on all states is $\rho^\sharp = \rho\otimes\tau$ where $\tau$ is some constant state of the bath.  This assignment map always lead to CP dynamics.  As such, one of his assumptions needs to be relaxed to find negative channels.  Most authors choose to give up the assumption of positivity on all reduced system states \cite{rodriguez10}.  Unfortunately, the concept of positivity domains seems ``unnatural'' to many physicists.  This topic will be addressed in more detail at the end of this paper.  For now, it suffices to point out that sharp operators with positivity domains are the only tools available to investigate negative channels unless one is willing to give up linearity or consistency.  

\section{Complete Positivity}
A positive map $\varepsilon$ is CP if $\varepsilon\otimes\mathbf{I}_n\ge0\;\forall n$ (where $\mathbf{I}_n$ is the $n$ dimensional identity operator).  There are two tests to check if a given map is CP: If $\varepsilon$ is CP, then is has an ``operator sum representation'', i.e.\
\begin{equation}
\varepsilon\otimes\mathbf{I}_n\ge0\;\forall n\Leftrightarrow \varepsilon\left(\rho\right) = \sum_i A_i \rho A_i^\dagger\;\;,
\end{equation}
where $A_i$ is an operator on the reduced system obeying certain requirements \cite{breuer07}.  The second test employs a specially constructed matrix as follows
\begin{equation}
\varepsilon\otimes\mathbf{I}_n\ge0\;\forall n\Leftrightarrow \mathbf{C} = \sum_{ij} E_{ij}\otimes \varepsilon\left(E_{ij}\right) \ge 0\;\;,
\end{equation}  
where $E_{ij}$ is a matrix with the same dimensions as the reduced system that has a 1 at the $ij$th position and 0 everywhere else.  The matrix $\mathbf{C}$ is commonly called ``Choi's matrix'' \cite{Choi75}.  These two tests are closely related \cite{breuer07}, but the second test leads directly to the definition of negativity for a channel.  

The negativity is defined as 
\begin{equation}
\eta \equiv \frac{\sum_i |\lambda_i|}{\sum_j |\lambda_j|} = \frac{1}{2}\left(1-\frac{\trace\left(\mathbf{C}\right)}{||\mathbf{C}||_1}\right)\;\;,
\end{equation} 
where $\lambda$ is an eigenvalue of $\mathbf{C}$, $\lambda_i<0\;\forall i$, and $||\mathbf{C}||_1$ is the trace norm of $\mathbf{C}$.  Notice, $\sum_j |\lambda_j|=\trace\left(\mathbf{C}\right)$ if and only if the negativity is zero.  From the definition, it is clear that $0\le \eta<\frac{1}{2}$ and
\begin{equation}
\varepsilon\otimes\mathbf{I}_n\ge 0\;\forall n\Leftrightarrow \eta = 0\;\;,
\end{equation}
i.e.\ a vanishing negativity implies CP.

Suppose the system of interest is a single qubit (i.e.\ a two-level quantum system).  In the single qubit case, Choi's matrix takes a simple block form, i.e.
\begin{equation}
\label{eq:Choi}
\mathbf{C} = \left(\begin{array}{c|c}
\varepsilon\left(\ketbra{0}{0}\right)&\varepsilon\left(\ketbra{0}{1}\right)\\
\hline
\varepsilon\left(\ketbra{1}{0}\right)&\varepsilon\left(\ketbra{1}{1}\right)\\
\end{array}\right)\;\;.
\end{equation}
The assumed linearity of the channel allows the off-diagonal blocks to be found using single qubit process tomography \cite{nielsen00}; hence, there is a connection between $\mathbf{C}$ and the tomographic characterization of a channel.  Eqn.\ \ref{eq:Choi} allows for quick determination of the negativity of channel.  For example, suppose a channel takes every input state to the completely mixed stated.  From Eqn.\ \ref{eq:Choi} it can be seen that this channel will have a negativity $\eta=0$.

It should be recognized that the mathematical definition of CP can have a clear physical interpretation.  Suppose $\rho$ is the initial state of some finite dimensional bipartite system associated to the Hilbert space $\mathcal{H}^{SB}=\mathcal{H}^S\otimes\mathcal{H}^B$.  Consider the Choi representation of a single qubit channel $\varepsilon$ with a single qubit bath written as
\begin{equation}
\mathbf{C} = \sum_{i,j=0}^1\ketbra{i}{j}\otimes\varepsilon\left(\ketbra{i}{j}\right) = I\otimes\varepsilon\left(\sum_{i,j=0}^1\ketbra{ii}{jj}\right)\;\;,
\end{equation}
where $I$ is the single qubit identity operator.  This form of Choi's matrix has led to the interpretation of Choi's matrix as a map $\varepsilon$ acting on one part of an (unnormalized) maximally entangled pair.  Define a state in the composite space as
\begin{equation}
\rho^\prime = I\otimes\varepsilon\left(\rho\right)\;\;.
\end{equation}
It follows that
\begin{equation}
\rho = N\sum_{i,j=0}^1\ketbra{ii}{jj} \rightarrow \rho^\prime = N\mathbf{C}\;\;,
\end{equation}
where $N$ is the appropriate normalization factor.  If $\rho^\prime$ must be a valid quantum state for any $\rho$, then $\rho^\prime\ge 0$ which implies $\mathbf{C}\ge 0$.  Thus, $\varepsilon$ must be CP.  This is the total domain argument for CP (which will be discussed again below), and it leads to the interpretation of CP as a requirement due to possible entanglement between bipartite subsystems.

\section{Examples of Negative Dynamics}
Imagine a two qubit universe as seen in Fig.\ \ref{fig:diagram}.  The channel would be the reduced dynamics of one of the qubits and would be described by some $\mathbf{C}$.  Given a time independent composite Hamiltonian $\hat{H}$, the evolution in the energy basis (i.e.\ the eigenbasis of $\hat{H}$) is described by a diagonal operator $U = \operatorname{diag}(e^{-i\nu_1t},e^{-i\nu_2t},e^{-i\nu_3t},e^{-i\nu_4t})$ where $\{\nu_i\}$ is an eigenvalue of $\hat{H}$ and, for convenience, everything is in units of $\hbar=1$.  The composite evolution is
\begin{equation}
\rho^{\sharp}_{ij}(t) = \left(U\rho^{\sharp} U^\dagger\right)_{ij} =  e^{-i(\nu_i-\nu_j)t}\rho^{\sharp}_{ij}\;\;.
\end{equation}
 
A sharp operator can be written in terms of a canonical tomographic basis 
\begin{equation}
\vec{\tau} =\{\ketbra{0}{0},\ketbra{+}{+},\ketbra{+_i}{+_i},\ketbra{1}{1}\}
\end{equation}
where $\ket{+} = 2^{-1/2}(\ket{0}+\ket{1})$ and $\ket{+_i} = 2^{-1/2}(\ket{0}+i\ket{1})$, as
\begin{equation}
\label{eq:Ass1}
\tau_{i}^\sharp = \tau_{i}\otimes\tau_{i}\;\;.
\end{equation}
The composite dynamics would then be described for the $\vec{\tau}$ basis with $\rho=\{\vec{\tau}\}_i\equiv \tau_i$ for $i=1,2,3,4$.  This sharp operator, along with the composite evolution $U$, would yield a channel described by
\begin{equation}
\mathbf{C} = \begin{pmatrix}
1&0&0&z\\
0&0&0&0\\
0&0&0&0\\
z^*&0&0&1\\
\end{pmatrix}\;\;,
\end{equation}
with $z=(1/2)(e^{-i(\nu_1-\nu_3)t}+e^{-i(\nu_2-\nu_4)t})$.  This situation leads to $zz^*=\cos^2(f_\nu t/2)$, where $f_\nu = \nu_1-\nu_2-\nu_3+\nu_4$ is a function of the Hamiltonian eigenvalues, and this result, in turn, implies the only two non-zero eigenvalues of $\mathbf{C}$, i.e.\ $1-\sqrt{zz^*}$ and $1+\sqrt{zz^*}$, are always positive and bounded $\in [0,2]$.  Hence, $\mathbf{C}$ always has a negativity $\eta=0$.  This sharp operator always leads to CP dynamics in the energy basis for any $\hat{H}$.  

It might be argued that the CP dynamics are a consequence of assigning the initial reduced state $\rho$ to a composite state with no entanglement (e.g.\ the concurrence $\mathcal{C}$ yields $\mathcal{C}(\rho^\sharp)=0\;\forall\rho$)\cite{Wootters01}.  This conjecture, however, can be proven false by counter example:  The sharp operator
\begin{equation}
\label{eq:Ass2}
\tau_{i}^\sharp = \tau_{i}\otimes(H\tau_{i}H^\dagger)\;\;,
\end{equation} 
where $H$ is the Hadamaard operator \cite{nielsen00}, also has no entanglement, yet it leads to
\begin{equation}
\label{eq:C2}
\mathbf{C}^\prime = \begin{pmatrix}
1&0&0&m\\
0&0&n&0\\
0&n^*&0&0\\
m^*&0&0&1\\
\end{pmatrix}\;\;,
\end{equation}
where $m=(1/4)(3e^{-i(\nu_1-\nu_3)t}+e^{-i(\nu_2-\nu_4)t})$ and $n=(1/4)(e^{-i(\nu_3-\nu_1)t}-e^{-i(\nu_4-\nu_2)t})$.  The four eigenvalues of $\mathbf{C}^\prime$ are
\begin{eqnarray}
1-\sqrt{mm^*} &=& 1-\sqrt{\frac{1}{8}\left(5+3\cos\left(f_\nu t\right)\right)}\;\;,\\
1+\sqrt{mm^*} &=& 1+\sqrt{\frac{1}{8}\left(5+3\cos\left(f_\nu t\right)\right)}\;\;,\\
-\sqrt{nn^*} &=& -\frac{\sin\left(f_\nu t/2\right)}{2}\;\;,\mathrm{and}\\
\sqrt{nn^*} &=& \frac{\sin\left(f_\nu t/2\right)}{2}\;\;.
\end{eqnarray}
Hence $\mathbf{C}^\prime$ will have $\eta>0$ unless $f_\nu t = 2n\pi$ where $n$ is some integer.  Notice $f_\nu$ is a function of $\hat{H}$ alone.  $\hat{H}$ determines the negativity of this channel independent of the lack of entanglement in the initial composite state.

\begin{figure} [t]
\centering
\includegraphics[width=0.5\textwidth]{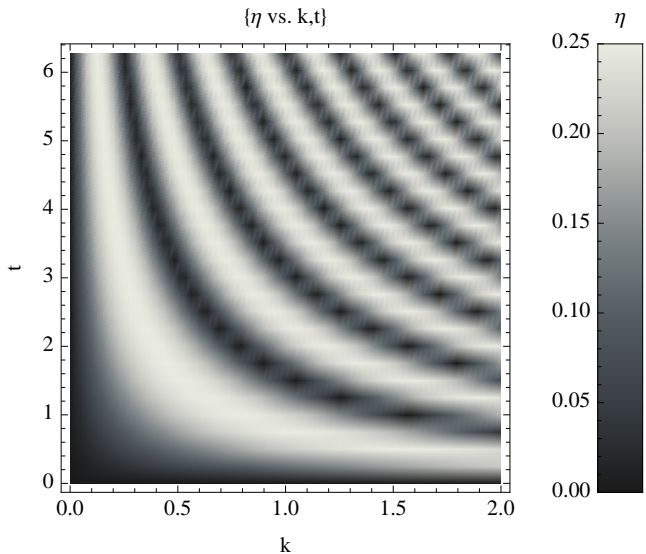}
\caption{Plot of the negativity of the channel defined by Eqn.\ \ref{eq:Ham1} and Eqn.\ \ref{eq:Ass2} as a function of time $t$ and the system-bath coupling $k$.}
\label{fig:IsingPlot}
\end{figure}

Our two qubit universe might be represented by a simple Hamiltonian, e.g.\
\begin{equation}
\label{eq:Ham1}
\hat{H} = \frac{\sigma_3}{2}\otimes\sigma_0 + \sigma_0\otimes\frac{\sigma_3}{2} + k\sigma_3\otimes\sigma_3\;\;,
\end{equation} 
where $\{\sigma_0,\sigma_1,\sigma_2,\sigma_3\}$ are the standard Pauli operators and $k$ is some coupling constant.  This Hamiltonian leads to $f_\nu = 4k$, which implies $\eta=0$ only when $kt=n\pi/2$ for the channel described by $\mathbf{C}^\prime$.  Fig.\  \ref{fig:IsingPlot} shows the negativity of this channel as a function of time and the coupling constant.

Eqn.\ \ref{eq:Ham1} has no special properties that make it a ``negative Hamiltonian''.  For example, $\hat{H}^\prime=\hat{H}+k^\prime\sigma_1\otimes\sigma_1$ yields $f_\nu = 2(-k^\prime+\sqrt{1+k^{\prime 2}})$, which implies $\eta=0\Rightarrow k^\prime = (-n^2 \pi^2 + t^2)/(2 n \pi t)$.  These Hamiltonians guarantee neither $\eta=0$ or $\eta>0$.  The negativity of the channel is a function of {\it both} the initial composite state and the composite Hamiltonian.  Given $\sharp$ defined by Eqn.\ \ref{eq:Ass2}, any Hamiltonian almost always leads to a negative channel in the energy basis.  Such a channel would only be completely positive, as pointed out above, if $f_\nu t = 2n\pi$.

\section{Discord of Initial Composite State}
It was recently shown that zero discord initial states always lead to CP reduced dynamics \cite{Rodriguez2008, Rodríguez07, Lidar2009a}, and these results might appear to suggest that Eqn.\ \ref{eq:Ass1} always describes a zero discord composite state (since it leads to CP reduced dynamics in the example) and that Eqn.\ \ref{eq:Ass2} never describes a zero discord composite state (since it leads to negative reduced dynamics in the example).  

The known results about guaranteed CP reduced dynamics are as follows:
\begin{itemize}
\item Regardless of initial correlations (i.e.\ for any initial composite state), local unitary composite evolution always leads to completely positive reduced dynamics \cite{Hayashi2003}.
\item Regardless of coupling (i.e.\ for any composite system evolution), zero discord initial composite states always lead to completely positive reduced dynamics \cite{Lidar2009a}.
\end{itemize}

None of the composite dynamics used in this paper are of the local unitary form, so the zero discord result seems like the result that might be applicable.  Notice, however, that Eqn.\ \ref{eq:Ass1} acting on the canonical tomography vector can lead to reduced dynamics with a negativity of $\approx 0.23$ if the composite dynamics are described by a controlled NOT gate (i.e.\ $\operatorname{CX}$ \cite{nielsen00}).  Similarly, Eqn.\ \ref{eq:Ass2} acting on the canonical tomography vector can lead to CP reduced dynamics given composite dynamics described by $\operatorname{CX}$.  So, neither example sharp operation leads to CP dynamics independently of the composite dynamics.  The negativity in the previous examples is a function of both the correlation and coupling in the system and can not be directly related to the discord in the initial composite state.

At this point in the discussion, a quick note should be made about the zero discord result.  A state with only classical correlations between the subsystems of $\mathcal{H}^X$ and $\mathcal{H}^Y$ can be written in the form \cite{Rodriguez2008}
\begin{equation*}
\rho^{XY} = \sum_i \left(\Pi^X_i\otimes I\right) \rho^{XY} \left(\Pi^X_i\otimes I\right)\;\;,
\end{equation*}
where $\Pi^X$ is a projector in the Hilbert space $\mathcal{H}^X$.  This result implies that a zero discord initial composite state would be in the form \cite{Rodriguez2008}
\begin{equation*}
\left(\rho^{S}\right)^\sharp = \sum_i \left( \Pi^S_i \otimes I\right) \rho^{SB} \left(\Pi^S_i \otimes I \right) = \sum_i \lambda_i \Pi^S_i \otimes \rho^B_i\;\;,
\end{equation*}
with $\lambda_i\ge 0\;\forall i$, $\sum_i \lambda_i = 1$, $\Pi^S_i=\ketbra{s_i}{s_i}$ where \{$\ket{s_i}$\} is an orthonormal basis of $\mathcal{H}^S$, and $\rho^B_i$ is a valid density operator in $\mathcal{H}^B$.  This initial composite state leads to an initial reduced state of
\begin{eqnarray}
\rho^S &=& \trace_B\left(\left(\rho^{S}\right)^\sharp\right)\\
&=&\trace_B\left(\sum_i \lambda_i \Pi^S_i \otimes \rho^B_i\right)\\
&=& \sum_i \lambda_i \Pi_i^S\;\;.
\label{eqn:zdrs}
\end{eqnarray}

The discord of a quantum state is basis dependent in the sense that it depends on the specific projectors $\Pi^S_i$.  The reduced dynamics are always CP for any composite dynamics if the reduced system can be written in the form of Eqn.\ \ref{eqn:zdrs}; i.e.\ the CP of the reduced dynamics in this proof require that the reduced state system be a convex sum of a given complete set of projectors $\{\Pi_i\}$.  In particular, this limitation means that the zero discord result is not very useful for channels defined on a tomography basis.

To see this point, notice that a tomography basis $\vec{\tau}^{\;\sharp}$ might consist of composite states that have zero discord with respect to different projectors, but the reduced dynamics associated to $\vec{\tau}^{\;\sharp}$ will only be completely positive if every reduced state in $\vec{\tau}$ could be written in the zero discord form of Eqn.\ \ref{eqn:zdrs} with respect to the same set of projectors $\{\Pi_i\}$.  However, if every reduced state in $\vec{\tau}$ could be written in this way, then $\vec{\tau}$ would not be a tomography basis.  For example, consider the states of the canonical qubit tomography vector.  Neither $\ketbra{+}{+}$ nor $\ketbra{+_i}{+_i}$ can be written in a zero discord form using $\{\ketbra{0}{0},\ketbra{1}{1}\}$, and similar troubles arise trying to use $\{\ketbra{+}{+},\ketbra{-}{-}\}$ or $\{\ketbra{+_i}{+_i},\ketbra{-_i}{-_i}\}$ as the projector sets for a zero discord form of the states in $\vec{\tau}$.  The Choi representation of the reduced dynamics is the ``workhorse'' representation in calculating the negativity, and the Choi representation comes from process tomography.  As such, the zero discord result does not give much insight into the origin of negativity in process tomography experiments.

\section{Physical Sharp Operations}
The above examples of negative channels depend on both the composite dynamics and the sharp operation.  The composite dynamics all are familiar, and, as such, they require very little physical motivation.  The sharp operator, however, is not so familiar.  An important question is whether or not the sharp operator of Eqn.\ \ref{eq:Ass2} physically reasonable.

Suppose the composite system is initially in the state $\ket{\Psi} = N(\ket{00}+\ket{01}+\ket{10}-\ket{11})$ where $N$ is the appropriate normalization factor.  This state, for example, might be the equilibrium state of our example universe.  Now suppose preparation of the reduced system is done with a perfect measurement procedure, i.e.\ to prepare the reduced system state as $\ket{0}$, the experimenter applies a projective measurement
\begin{equation}
\left(\ketbra{0}{0}\otimes I\right)\ket{\Psi} = \ket{0+} = \left(I\otimes H\right)\ket{00}\;\;.
\end{equation}
Similarly, measurement-preparation of the other three tomographic basis states yield initial composite states of $\left(I\otimes H\right)\ket{11}$, $\left(I\otimes H\right)\ket{++}$, and $\left(I\otimes H\right)\ket{+_i+_i}$.  Hence, the $\sharp$ operation from Eqn.\ \ref{eq:Ass2} can be thought of as a measurement-preparation procedure on a composite system that is initially in a superposition state $\ket{\Psi}$.  

It has recently been shown that preparation of the reduced system by a projective measurement on an initially entangled composite system state leads to negative reduced dynamics in situations beyond the examples given here \cite{Devi2011}.  The physical interpretation of such sharp operations is that of an ideal preparation of a system that is initially entangled with the bath.  Such sharp operations can be produced in the lab through the use of ``controlled bath'' type experiments.

The sharp operator is linear and consistent by definition.  However, $\sharp$ is not defined on the space of all possible composite states.  Notice, $(\ketbra{-}{-})^\sharp$, where $\ketbra{-}{-} = \tau_3+\tau_4-\tau_1$, is not a valid state when $\sharp$ is defined by Eqn. \ref{eq:Ass2}; i.e.\ $(\ketbra{-}{-})^\sharp < 0$.  The $\sharp$ operator is positive on the tomographic basis states $\vec{\tau}$, which are the only reduced states ever prepared in the lab, but the fact that it can lead to non-positive states (e.g.\ states with negative occupation probabilities) makes some authors nervous.

Following this line of logic, some authors impose the requirement of complete positivity as an empirical requirement using what is termed the ``total domain argument'' for CP.  The total domain argument proceeds as follows: The density matrix of a reduced system must be positive by definition, and it is clear that no system is ever truly isolated in the lab.  The density matrix describing the reduced system in contact with a non-interacting environment must still be positive.  If a map is a valid quantum map, then it must take valid density matrices to valid density matrices.  A trivial extension of the map is physically reasonable and must result in a valid quantum map, i.e.\ the trivial extension of the quantum map must also take valid density matrices to valid density matrices.  Therefore, the quantum map must be CP.  

The total domain argument can be avoided through the use of positivity domains (which have already been introduced).  Notice that the positivity domain is the domain of states in which a map $\Gamma$ will be positive.  On the positivity domain, $\Gamma$ will take valid initial states to valid final states.  Such a requirement is identical in spirit to the total domain argument for the CP requirement, except that it is not extended to states which are not actually created in the lab.

The reduced dynamics are defined by the composite dynamics, the tomography basis, and the sharp operation.  Notice that if the reduced dynamics have a non-trivial positivity domain, then that domain will depend on the tomography basis used to form the Choi representation.  Each different tomography basis used to form a Choi representation will have a different positivity domain.  Hence, demanding CP reduced dynamics is a demand of reduced dynamics that are independent of the tomography basis used to write them down.

The positivity requirement of $\sharp$ on the space of all composite states can be imposed as a ``physical'' (or philosophical) requirement, but it is not required mathematically.  

\section{Extended Baths}
It might be argued that the negativity arises from the simplicity of the example universe.  In general, if the reduced system is a qubit, the composite system consists of $M$ qubits, and the energy basis evolution is described by some time independent Hamiltonian $\hat{H}$, then given an assignment map defined as
\begin{equation}
\vec{\tau}^\sharp=\vec{\tau}\otimes \left(H\vec{\tau}H^\dagger\right)\otimes \bigotimes_{i=3}^M \tau_3\;\;,
\end{equation}
this qubit channel would be described by
\begin{equation}
\label{eq:C3}
\mathbf{C}^{\prime\prime} = \begin{pmatrix}
1&0&0&m^\prime\\
0&0&n^\prime&0\\
0&n^{\prime*}&0&0\\
m^{\prime*}&0&0&1\\
\end{pmatrix}\;\;,
\end{equation}
 with $m^\prime=(1/4)(3e^{-i(\nu_1-\nu_s)t}+e^{-i(\nu_{1+r}-\nu_{s+r})t})$ and $n=(1/4)(e^{-i(\nu_s-\nu_1)t}-e^{-i(\nu_{s+r}-\nu_{1+r})t})$ with $s=(2^M/2)+1$ and $r=2^M/4$.  Again, $\nu_i$ is an eigenvalue of $\hat{H}$.  $\mathbf{C}^{\prime\prime}$ will have $\eta>0$ unless $f^\prime_\nu t = 2n\pi$ where $n$ is some integer and $f^\prime_\nu = \nu_1-\nu_{1+r}-\nu_s+\nu_{s+r}$.  This M-qubit channel will be negative except for very specific values of $t$ and the Hamiltonian.  In the energy basis of the $\hat{H}$, this channel is almost always negative.  

\section{Experiments}
Consider, again, an example universe of two qubits.  One qubit will be the reduced system and the other will act as the bath.  If the composite dynamics are defined as the root swap gate
\begin{equation}
U_{\sqrt{Sw}} = \frac{1}{\sqrt{2}}\begin{pmatrix}
\sqrt{2}&0&0&0\\
0&1&i&0\\
0&i&1&0\\
0&0&0&\sqrt{2}
\end{pmatrix}\;\;,
\end{equation}
then the channel becomes
\begin{equation}
\varepsilon(\rho) = \trace_B\left(U_{\sqrt{Sw}} \rho^\sharp U_{\sqrt{Sw}}^\dagger\right)\;\;.
\end{equation} 
Define the sharp operation as Eqn.\ \ref{eq:Ass2} on the canonical tomography vector $\vec{\tau}$ (introduced above).  Process tomography of this channel yields a Choi representation of
\begin{equation}
\mathbf{C}_{\sqrt{Sw}} = \begin{pmatrix}
 \frac{3}{4} & -\frac{i}{2 \sqrt{2}} & \frac{1}{4} & \frac{\frac{1}{2}+\frac{i}{2}}{\sqrt{2}} \\
 \frac{i}{2 \sqrt{2}} & \frac{1}{4} & \frac{\frac{1}{2}-\frac{i}{2}}{\sqrt{2}} & -\frac{1}{4} \\
 \frac{1}{4} & \frac{\frac{1}{2}+\frac{i}{2}}{\sqrt{2}} & \frac{1}{4} & -\frac{i}{2 \sqrt{2}} \\
 \frac{\frac{1}{2}-\frac{i}{2}}{\sqrt{2}} & -\frac{1}{4} & \frac{i}{2 \sqrt{2}} & \frac{3}{4}
\end{pmatrix}\;\;,
\end{equation}
and a channel negativity of $\eta_{\sqrt{Sw}} \approx 0.149$.

The root-swap gate has been accomplished on polarization state photonic qubits \cite{Cernoch2008} and those experiments can be modified to experimentally test negativity calculations.  Notice, however, that the experiment would need to be modified further than simply skipping the maximum likelihood reconstruction to calculate the negativity.  The negative channel here is a single qubit channel, not the two qubit channel that was tomographically characterized in \cite{Cernoch2008}.  But, this experiment can be modified by adding a Hadamard rotation (to the polarization state) in one of the two arms in the preparation phase of the set-up.  Performing {\em single} qubit process tomography on one of the two qubits in this experiment would lead to a superoperator (or Choi) representation from which the negativity could be measured.

If the composite dynamics are defined as a controlled phase gate
\begin{equation}
CZ = \begin{pmatrix}
1&0&0&0\\
0&1&0&0\\
0&0&1&0\\
0&0&0&-1
\end{pmatrix}
\end{equation}
then the channel becomes
\begin{equation}
\varepsilon(\rho) = \trace_B\left(CZ \rho^\sharp CZ^\dagger\right)\;\;.
\end{equation} 
A Choi representation of this channel can be found using the sharp operator from the previous example; i.e.\
\begin{equation}
\label{eqn:Ccz}
\mathbf{C}_{CZ} = \begin{pmatrix}
 \frac{1}{2} & \frac{1}{2} & \frac{1}{2} & -\frac{1}{2}-\frac{i}{2} \\
 \frac{1}{2} & \frac{1}{2} & -\frac{1}{2}-\frac{i}{2} & -\frac{1}{2} \\
 \frac{1}{2} & -\frac{1}{2}+\frac{i}{2} & \frac{1}{2} & \frac{1}{2} \\
 -\frac{1}{2}+\frac{i}{2} & -\frac{1}{2} & \frac{1}{2} & \frac{1}{2}
\end{pmatrix}\;\;.
\end{equation}
The negativity of this channel is $\eta_{CZ}\approx 0.167$.

Again, the complete experiment would involve the preparation of the tomography vector and sharp operation, and then the application of the $CZ$ gate.  The desired sharp operation is straightforward to implement in an optical set-up and can be done in exactly the same manner as described in the previous subsection: a non-linear crystal can be used to create an entangled pair of qubits, one of which is passed through a waveplate yielding a Hadamard rotation, the other of which is prepared with polarization filters.  Encoding the qubit in the polarization of the photons means all of the desired operations can be accomplished with well understood polarization optics.  

These experiments provide exactly the desired ``controlled bath'' situation needed to experimentally study and verify the theoretical predictions of negativity.  The sharp operation can be changed by interchanging the Hadamard rotation in the preparation stage of the set-up with some other rotation.  In this way, the theoretical predications concerning the impact of the sharp operation can be tested experimentally.  

\section{Conclusions}
Negative channels appear to be physically realizable.  The CP requirement on quantum channels imposes restrictions on the mathematical representations of channels, which in turn, might be stifling some of their possible utility.  More importantly, the CP requirement might be leading experimenters to ignore data that might otherwise give important clues into faulty preparation procedures or system-bath coupling.  For example, Cory et al.\ \cite{Cory2004} performed process tomography on an NMR system and found a channel with $\eta\approx 0.29$.  The CP requirement also has physical consequences beyond tomographic channel characterization which also seem to be contradicted by experimental evidence, as pointed out by other authors \cite{Pechukas1,benatti03}.  

The question of whether or not ``physical'' reduced dynamics need to be completely positive is still an open question.  But, it is possible to write down theoretical descriptions of tomography experiments which lead to reduced dynamics that appear negative, and tomography experiments can be conducted in the lab using ``controlled baths'' that similarly yield reduced dynamics that appear negative.  Examples of both situations are given in this article.

A channel with a vanishing negativity is mathematically convenient.  Mathematical convenience, however, is not a reason to exclude possibly relevant experimental data.  It is important to understand the apparent negativity of the reduced dynamics presented in this article, as well as the non-zero negativities that have been measured in various tomography experiments.  The negativity of the Cory et al.\ experiment mentioned above has been investigated to determine if it could be explained entirely by statistical errors associated with the tomography process \cite{Wood2009}, and that author determined that statistical errors could not account for all apparently negative channels.  Notice that statistical error was not discussed in any of the examples presented in this article.  For example, all of the measurements used the tomography processes described above were assumed to be perfect.  None of the non-zero negativities presented here can be explained as ``experimental'' (or ``statistical'') error.

Given the ubiquity of the CP assumption, and its importance in many quantum information theories, it is important to understand how and when the CP assumption is justified.  If the CP assumption is always justified, then the non-zero negativities seen in the examples above must be an artifact of some ``un-physical'' part of the proposed experiments.  Notice, however, that the proposed ``controlled bath'' experiments allow the measurement of non-zero negativities in straightforward process tomography experiments that are already a common part of quantum information theory.  The sharp operations likewise have straightforward physical implementations.  As such, if these descriptions are ``un-physical'', then that is a subtle point that needs to be better understood.

I'd like to thank Keye Martin, Marco Lanzagorta, Johnny Feng, and Tanner Crowder for their help with this idea.  

\bibliographystyle{apsrev4-1}
\bibliography{negativity}
 
\end{document}